\documentclass[preprint,aps,eqsecnum,nofootinbib]{revtex4}
\usepackage{epsfig}

\newcommand{\be}{\begin{equation}}
\newcommand{\ee}{\end{equation}}
\newcommand{\bea}{\begin{eqnarray}}
\newcommand{\beas}{\begin{eqnarray*}}
\newcommand{\eea}{\end{eqnarray}}
\newcommand{\eeas}{\end{eqnarray*}}
\newcommand{\ba}{\begin{array}}
\newcommand{\ea}{\end{array}}
\def\ls{\mathrel{\lower4pt\vbox{\lineskip=0pt\baselineskip=0pt
           \hbox{$<$}\hbox{$\sim$}}}}
\def\gs{\mathrel{\lower4pt\vbox{\lineskip=0pt\baselineskip=0pt
           \hbox{$>$}\hbox{$\sim$}}}}

\begin{document}


\preprint{HIP-2004-08/TH}

\title{Dumping inflaton energy density out of this world}
\author{Kari Enqvist~$^{1}$, Anupam Mazumdar~$^{2}$, and
A. P\'erez-Lorenzana~$^{3}$}

\affiliation{$^{1}$~Department of Physical Sciences, University of Helsinki,
and Helsinki Institute of Physics, P.O. Box 9, FIN-00014 University of
Helsinki, Finland\\
$^{2}$~CHEP, McGill University, 3600 University Road, Montr\'eal,
Qu\'ebec, H3A 2T8, Canada\\
$^{3}$~Departamento de F\'{\i}sica, Centro de Investigaci\'on y de Estudios
Avanzados del I.P.N.\\
 Apdo. Post. 14-740, 07000, M\'exico, D.F., M\'exico}


\begin{abstract}
We argue that a brane world with a warped, infinite extra
dimension allows for the inflaton to decay into the bulk so that
after inflation, the effective dark energy disappears from our
brane. This is achieved by the redshifting of the decay products
into infinity of the 5th dimension. As a consequence, all matter
and CMB density perturbations could have their origin in the decay
of a MSSM flat direction rather than the inflaton. We also discuss
a string theoretical model where reheating after inflation may not
affect the observable brane.
\end{abstract}

\maketitle
\section{Introduction}

Recent observations~\cite{WMAP} strongly support a period of
primordial inflation. Besides making the universe flat and
homogeneous, inflation is the only known dynamical mechanism which can
stretch small quantum fluctuations outside the Hubble horizon.  These
perturbations act as seeds for the large scale structures in the
Universe. However, despite of the success of the inflationary paradigm
(for a review, see ~\cite{Riotto}), we know very little about the
inflationary sector. The inflaton potential must nevertheless be very
flat with a very small self coupling; likewise, its coupling to other
fields must be extremely weak.  Such small couplings are hard to come
by without fine tuning, which renders inflaton to a gauge singlet
whose couplings can be adjusted at our will. Given this, the immediate
question is, what is the inflaton decaying into?

Eventually the cold inflationary Universe must reheat with the
Standard Model (SM) degrees of freedom, or, as the current theoretical
prejudice dictates, with the Minimally Supersymmetric Standard Model
(MSSM) degrees of freedom. MSSM contains a number of flat directions
along which the renormalizable part of the potential vanishes (for a
review, see~\cite{Enqvist}). During inflation the massless fields
corresponding to the flat directions receive scale-invariant
perturbations. It would be tempting to associate them with reheating
and the generation of CMB perturbations
\cite{Kasuya2,Kasuya3,Kasuya1,Others}.  This would require that the
flat directions will eventually dominate over the radiation produced
by the inflaton decay. In other words, the flat direction would have
to act as an MSSM curvaton~\cite{Sloth,Lyth,Moroi}. However, it turns
out that such domination is not possible unless the inflaton decays
into some hidden degrees of freedom, rather than into MSSM radiation
\cite{Kasuya2,Kasuya3}.

In this respect brane world scenarios, where the Universe is
regarded as a three dimensional hypersurface embedded in a higher
dimensional bulk, bring along new, interesting possibilities. The
SM degrees of freedom are assumed to be stuck on a brane while
gravity is propagating in the entire bulk. The bulk could also
have a non-trivial background geometry which allows for the zero
mode of graviton to be trapped at the brane location, such as in
the case of an anti de-Sitter (adS) bulk in the Randall-Sundrum
type models~\cite{RS1,RS2}. The value of the Newton's constant
requires the fundamental scale to be fairly large, $10^{18}~{\rm
GeV}\geq M_{s}\geq 10^{3}$~GeV. In these models the inflaton,
again treated as a gauge singlet, could either live on the brane
or in the entire bulk. There is also the exciting possibility that
the inflaton energy density does not reheat the Universe, but gets
deposited in the infinite bulk or onto the adS horizon. In this
respect the inflaton potential could be treated as a kind of dark
energy for the MSSM brane. It has been shown \cite{Kasuya1} that
if the dark energy of the inflaton field can be transferred into
the bulk so that it no longer is visible on our brane, even the
simplest MSSM flat directions can act as curvatons and provide
both all the matter and the density perturbations in the Universe.

The brane world scenarios are supported by the string theory,
where there is a natural explanation for the construction of MSSM
like brane with a help of coincident stack of $Dp$ branes ($p$ is
the number of spatial dimensions along the brane, $p$ is odd
(even) in type IIA (IIB) string theories) attached to some
orbifold point~\cite{Quevedo1}~\footnote{For different
constructions, see~\cite{Quevedo3}.}. The open strings attached to
$Dp$ branes act as sources for gauge fields and gravity again
propagates in the entire bulk. In this kind of framework the
inflaton may be regarded as a moduli, for example the physical
separation between a $Dp$ and anti-brane $\bar Dp$ brane, or the
angular separation between $Dp-Dp$ or $Dp-\bar Dp$ branes (for a
review see~\cite{Quevedo2}).  Inflation may end by virtue of
tachyon condensation when branes approach close to the string
scale~\cite{Sen}, or with a help of many tachyons as in the case
of assisted inflation~\cite{Panda}. It is however not guaranteed
that inflaton will reheat the Universe with the MS(SM) degrees of
freedom. One could rather argue that it is more likely that the
inflaton will reheat the bulk.

The purpose of this paper is to construct a brane world model with
a warped bulk so that it is possible to localize the inflaton
energy density away from the MS(SM) like branes. We will argue
that the inflaton energy can be redshifted away so that after
inflation there is effectively no energy density other than that
of the excited MSSM flat directions on our brane. For our
purposes, the inflaton potential is a form of dark energy which is
only responsible for making the Universe, parallel to the brane
directions, large, homogeneous and isotropic.

The paper is organized as follows. In Section II we recapitulate
some known results of the brane world models and discuss infinite
extra dimensions. In Section III we  present a brane-world where
the dark energy can be dumped into the bulk instead onto the brane
and estimate the escape rate of the inflaton decay products from
the brane. In Sect. IV we motivate the model by considerations
relating to string theoretical inflation. In Sect. V we give our
conclusions and discuss how the present results can be combined
with curvaton-like scenarios involving MSSM flat directions to
yield all matter and an adiabatic, scale-invariant spectrum of
perturbations.


\section{Infinite extra dimension and KK decomposition}

In this section we briefly recapitulate some of the already known
results of the brane world models.  We start with the simplest
scenario assuming that there is a three dimensional hypersurface,
called the brane, which carries MSSM degrees of freedom. The MSSM
brane is embedded in a $5$ dimensional space (the bulk) with a
non-factorizable metric \cite{RS1,RS2}~(for a nice review,
see~\cite{Rubakov1})
\begin{equation}
\label{metric}
ds^2=a^2(z)\eta_{\mu\nu}dx^{\mu}dx^{\nu}-dz^2\,,
\end{equation}
where $\eta_{\mu\nu}$ is the four dimensional Minkowski metric. We
take the  extra dimension to be  infinite. The  brane is located
at $z=0$. The total action for gravity is given by
\begin{equation}
S_{g}=-\frac{1}{16\pi G_{5}}\int d^4x dz\sqrt{g^{(5)}}R^{(5)}-\Lambda
\int d^{4}xdz\sqrt{g^{(5)}}-\sigma\int d^{4}x\sqrt{g^{(4)}}\,.
\end{equation}
where $\Lambda$ is a bulk negative cosmological constant which is
related to the brane tension $\sigma$ by the fine tuning relationship
\begin{equation}
\Lambda=-\frac{4\pi}{3}G_{5}\sigma^2\,.
\end{equation}
The warp factor present in the metric has the form $a(z)=e^{-k|z|}$,
with
\begin{equation}
k=\frac{4\pi}{3}G_{5}\sigma\,.
\end{equation}
Here $G_{5}$ stands for the true gravitational coupling constant
of the five dimensional theory, which defines the fundamental
gravity scale as $M_{\ast}=(8\pi G_{5})^{-1/3}$. The effective
four dimensional Newton's constant $M_{P}=(8\pi
G_{4})^{-1/2}=2.4\times 10^{18}$~GeV is given by
\begin{equation}
G_{4}=k~G_{5}\,.
\end{equation}

This setup is inspired by adS/CFT correspondence~\cite{Maldacena},
where the entire bulk has adS geometry. It has an interesting
feature along the $z$ axis in that the fifth dimension has a
horizon at $z=\infty$. A particle that escapes from the brane and
moves along a geodesic travels from $z=0$ to $z=\infty$ in a
finite proper time $\tau=\pi/2k$. In other words, $z=\infty$ is a
particle horizon.

Another interesting property is that the spectrum of the KK
gravitons has a localized massless mode around the brane,
identified as the four dimensional graviton,  plus a continuum of
modes with wave functions which are oscillatory at large $z$,
\begin{equation}
h_{m}(z)= {\rm
const}~\times\sin\left(\frac{m}{k}e^{kz}+\phi_{m}\right)\,,
\end{equation}
while  near the brane location, at $z=0$, the wave function is
suppressed:
\begin{equation}
h_{m}(z=0)={\rm const}~\times\sqrt{\frac{m}{k}}\,.
\label{h0}
\end{equation}
By summing over all the KK modes, including the zero mode of the
graviton, one finds a modification of the gravitational
interaction between two point particles located on the brane. If
their  masses are $m_1$ and $m_2$ and the particle separation is
$r$, the potential has the form
\begin{equation}
V(r)=-\frac{G_{(4)}m_1 m_2}{r}\left(1+\frac{\rm const}{k^2r^2}\right)\,.
\end{equation}
For distances $r\gg 1/k$ the correction to the Newtonian gravity
is negligible small. Current experiments have tested the validity
of Newton's law down to sub-millimeter distances, which implies
that $k> 10^{-3}$~eV~\cite{expgrav}. In this paper we will mostly
assume that $k$ and $M_*$, are close to the four dimensional
Planck scale.

It is clear that with an infinite fifth dimension the KK modes of
other bulk fields also possess a continuum of modes~\cite{wise},
and may also have a quasi-localized mode~\cite{Rubakov2}. Consider
for instance the case of a scalar bulk field on the background
metric Eq.~(\ref{metric}). The action is then given by
 \begin{equation}
  S_\chi = \int\! d^4x\, dz\, \sqrt{g^{(5)}}
  \left({1\over 2} g^{ab} \partial_a\chi \partial_b\chi -
  {1\over2} \mu^2\chi^2 \right)~.
 \end{equation}
The corresponding KK wave function is then defined as a solution
to the field equation
 \be
  \left[-\partial_z^2 +
  4k\, {\it sgn}(z) \partial_z + \mu^2 - m^2/a^2(z)\right] \chi(z;m) =0~,
  \label{eqofmotion}
 \ee
where $m^2 = p^\mu p_\mu$ defines both the four dimensional mass
and the KK level. Eq. (\ref{eqofmotion}) is supplemented by the
boundary condition on the brane $\partial_z\chi(z=0;m) =0$, and
the normalization condition $\int dz\, a^2(z) \chi(z;m) \chi(z;m')
= k \delta(m-m')$. The general solution to Eq. (\ref{eqofmotion})
is given in terms of Bessel functions of index $\nu = \sqrt{4 +
\mu^2/k^2}$, and can be written as~\cite{wise,Rubakov2}
 \be
     \chi(z;m) = {1\over N(m) a^2(z)}\,
     \left[  J_\nu \left( {m\over ka(z)}\right) +
            A(m) Y_\nu \left( {m\over ka(z)}\right)
     \right]\,,
 \label{kkchi}
 \ee
where the normalization factor, $N$, and the coefficient $A$ are
functions of the continuous KK index $m$. One finds that
~\cite{wise,Rubakov2} $N(m) = \sqrt{1+A^2(m)}/\sqrt{m/2}$ with
 \be
    A(m) = -{2\, J_\nu(m/k) + (m/k) J'_\nu(m/k) \over
             2\, Y_\nu(m/k) + (m/k) Y'_\nu(m/k)}.
 \ee
For $\mu, m \ll k$ one can approximate the last expression by
taking $\nu=2$ and show that
 \be
 A(m)\approx {\pi\over 4} \left({m\over k}\right)^2~.
 \ee
Thus, the KK wave function evaluated at the brane is just
 \be
   \chi(0,m)\approx \sqrt{\frac{m}{2}}~.
 \label{chi0}
 \ee
Note that the functional behavior of the above expression is the
same as for the graviton case in Eq.~(\ref{h0}).

As noted in Ref.~\cite{Rubakov2}, this system has in general a
resonance around the brane, i.e., there is a quasi-localized mode
of non-zero mass; that is, there is no truly bound state in the
spectrum. This can be visualized in a simple way: the continuum of
modes is determined by the asymptotic form of the field equation
at large $z$, where the mass term $\mu^2$ is negligible compared
with $m^2/a^2(z)$, which shows that the spectrum does start at
$m=0$, independently of $\mu$, but there are no bound states
within the continuum. By exploring the KK modes one can show that
there is a mode that actually has a complex eigenvalue $m = m_0 +
i \Gamma$~\cite{Rubakov2}. Thus, this mode can be considered as a
quasi-localized metastable state for which $\Gamma$ gives the
escape rate from the brane into the extra dimension towards
infinity. For $\mu\ll k$ one finds $m_0 = \mu/\sqrt2$ and \be
\Gamma = {\pi\over 16}\left({ m_0\over k}\right)^2\, m_0~.
\label{gamma0} \ee In the following section we will use some of
these results when considering the decay life time of the
inflaton.


\section{Inflaton decay into the bulk in a warped brane world scenario}

Let us now discuss how after inflation the inflaton may disappear
from the brane and leave behind an (almost) empty brane, with a
minor impact on the later cosmological evolution. This is a
radical point of view that, however, can easily be accommodated
within the context of infinite extra dimension
models~\cite{Lorenzana}. To be more specific, let us assume that
the inflaton is a true 4D brane field, with a homogeneous
distribution that dominates the energy density at the early
Universe on the observable brane and gives rise to a period of
inflation. Then the Friedmann equation has a quadratic dependence
on the brane density $\rho$~\cite{langlois} so that \be H^2 =
{1\over 3 M_P} \rho \left(1 + {\rho\over 2\sigma} \right)\,,
\label{friedman} \ee which becomes the standard relation $H =
\sqrt{\rho/3 M_P}$ only for densities small compared to the brane
tension~\cite{langlois,carlos}.

Once inflation comes to an end, the inflaton will decay, but instead
of reheating the brane degrees of freedom, we now assume that it
couples to the bulk fields alone, and decays into bulk degrees of
freedom. This may happen, for instance, if the inflaton and the bulk
fields carry some global quantum number while the brane degrees of
freedom do not. All the inflaton energy would be radiated into the
empty bulk after the end of inflation in the form of KK modes. These
bulk modes carry momentum along the fifth dimension, so that they
would simply fly away into the empty bulk, towards infinity, taking
the inflaton energy away from the brane. The energy density of the
inflaton will be gradually redshifted into the bulk before becoming
vanishingly small. A small fraction, however, may act as a dark energy
on the brane. We will comment on this below.

It is interesting that whether the inflaton density is larger than
brane tension  or not becomes irrelevant for the purposes of the
present discussion. Inflation could well take place in the non
standard regime of the theory, without  leaving any visible trace
on the subsequent thermal evolution of the Universe~\cite{Maz}.

Our scenario can be thought of as a hot radiating plate cooling down
by emitting its energy into the cold surrounding space.  It is not
hard to see that such cooling process is extremely efficient. To
demonstrate the feasibility of the idea, let us consider the coupling
of the inflaton to a bulk scalar field $\varphi$, which in the
complete $5$D theory can be written as
\begin{equation}
 \sqrt{g(z)}\, h\, \phi(x)\, \varphi(x,z)\, \varphi(x,z)\,\delta(z)\,,
\end{equation}
with $h$ is corresponding coupling constant. After introducing the
KK decomposition of the bulk field and integrating out the extra
dimension one gets the effective coupling of the inflaton to the
KK modes as
 \be
 h\,[\chi(0,m) \chi(0,m')]~ \phi(x)\, \varphi_m(x)\, \varphi_{m'}(x)\,;
  \ee
where $\chi(0,m)$ are the $z$ dependent wave functions of a KK
modes of mass $m$, given in Eq.~(\ref{kkchi}), evaluated at the
brane position. The KK mass dependence of the effective couplings
indicate that the inflaton would preferably decay into the heavy
modes, i.e. those with the largest momentum along the fifth
dimension. This scenario is similar to the one discussed in
Ref.~\cite{Rouzbeh} for the cooling down of a hot brane by
graviton emission, although there the KK gravitons were assumed to
be thermally produced.

If the inflaton decays into the continuum of KK modes with masses
smaller than $m_\phi$, it is straightforward to estimate the total
decay rate as
 \be
   \Gamma_\phi = \int_{0}^{m_\phi}\int_{0}^{\sqrt{m_\phi^2 - m^2}}\,
            {dm\over k}{dm'\over k}\, h^2\,
        {[\chi(0,m) \chi(0,m')]^2 \over m_\phi}\,
     \approx
        {h^2\over 32} \left({m_\phi\over k}\right)^2\, m_\phi \,;
 \label{gamma}
 \ee
where the RHS has been estimated in the limit where $\mu,m_\phi\ll
k$  using Eq.~(\ref{chi0}). Since the inflaton is heavy, say,
about the GUT scale, whereas $k$ is close to $M_{P}$, the
suppression on the decay rate is not large. As a consequence the
inflaton may release all its energy into the bulk fields very
efficiently.

Let us now discuss what happens to the energy that has been
injected into the bulk. As already mentioned above, since KK modes
have a fifth momentum, they will travel away from the brane,
moving towards infinity. As the original energy density of the
inflaton field is shifted away from the brane into the extra
dimension, only the tail of the density distribution would be felt
by the brane. It has been pointed out, however, that the energy
emitted from the brane would eventually collapse to a black hole
at the end of the space~\cite{March-Russell}. The presence of the
black hole changes the brane expansion by  introducing a new
contribution to the Friedmann equation Eq.~(\ref{friedman}) which
behaves as ${2 M_{BH}/ a^4}$, where  $a$ is the brane scale factor
and $M_{BH}$ is a parameter interpreted as the 5D "mass" of the
black hole~\cite{visser}. This term acts as a dark energy, which,
provided that $M_{BH}$ is small, has a subleading role in the
early Universe. In addition, due to the warp factor of the metric
the whole energy density suffers a depletion as it moves through
the throat of the AdS space. This is the same phenomena that was
used to explain the hierarchy problem in the original
Randall-Sundrum papers~\cite{RS2}. As one moves away from the
brane, the mass scales shrink down; what is a large scale on the
brane at $z=0$ looks exponentially small at any other point on the
fifth dimension. This behavior helps to understand why the large
energy deposited into the bulk by the inflaton should appear small
once the decay products are hidden behind the black hole horizon.
Therefore, although we have not estimated the magnitude of the
black hole dark energy, we believe it should be harmless.

There are some alternatives to the picture presented here. For
example, one could use a bulk field to provide for the inflaton as
its quasi-localized (resonant) mode as described at the end of
previous section. It is interesting to note that the width of such
a mode, given by Eq.~(\ref{gamma0}), has the same functional form
as Eq.~(\ref{gamma}), so that its escape rate from the brane would
be as efficient as the  decay of the inflaton in the our present
model.

\section{A stringy motivation}

In string theories inflationary cosmologies have often been
discussed (see. e.g. ~\cite{Quevedo2}) by making use of the fact
that $Dp$ and $\bar Dp$ branes can attract each other by virtue of
the spacetime supersymmetry breaking~\cite{Polchinski}. If the
attractive potential is sufficiently flat then it can give rise to
a slow roll inflation~\cite{Burgess}. One of the constraints is
that the brane separation and hence the bulk has to be larger than
the inverse of the compactification scale, which is indeed a
problem \cite{Maldacena1} and can be viewed as an initial
condition problem requiring some kind of a bulk inflation
\cite{anupam} before brane inflation. Such early inflation could
be triggered, e.g. by gas of branes~\cite{Brandenberger1}.

An important issue is the stabilization of the volume, the dilaton
and the moduli. In the context of a warped background one
typically finds a Klebanov-Strassler kind of solution for the
background metric~\cite{Klebanov}. One can think of this geometry
as a stringy generalization of the Randall-Sundrum model, where
there is an adS throat (or a conical singular region) where the
infrared brane can be regularized by the infrared geometry. The
Klebanov-Strassler solution gives rise to a non-compact geometry
where the radial internal coordinate can be thought of as $z$ in
the $5$ dimensional language of Randall-Sundrum warped geometry.
The Planck brane is at the large $r$ ultraviolet region, while the
small $r$ infrared brane is stuck near the adS throat with, where
$r$ parameterizes the internal radial coordinate and the metric is
of the form
\begin{equation}
ds^2=h^{-1/2}(r)dx_{\mu}dx^{\mu}+h^{1/2}(r)(dr^2+r^2ds_{(5)})\,.
\end{equation}
A particular realization of such compactification on a Calabi-Yau
manifold gives rise to the stabilization of the complex structure
moduli~\cite{Joe1,Frey,Linde1}.

In this paper we will not be able to discuss a detailed
inflationary model within such stringy framework. We will merely
present some plausibility arguments supporting our case as
discussed in Sect. III . To this end, let us consider a scenario
which consists of a stack of branes mimicking MSSM gauge group
(stack of $D3$ branes and $D7-\bar D7$ branes (the latter ones
required for an exact cancellation of the tadpoles at the
singularity). In addition, there should be other branes and
anti-branes to drive inflation along the three spatial directions.
The MSSM branes are embedded near the ultraviolet part of the
geometry, at large $r$, far from the adS throat, while the branes
which give rise to inflation are moving towards the region where
the adS throat is located, near the infrared part of the geometry,
at small $r$.

To be concrete, let us assume that all the moduli are stabilized
like in~\cite{Linde1}. A set of $\bar D3$ branes are stuck near
the infrared part of the bulk, and inflation occurs because of the
slow motion of a $D3$ brane approaching from the ultraviolet
regime. The potential of $D3-\bar D3$ is also felt by the MSSM
branes and is described by a four dimensional effective field
theory on the world volume. The presence of $\bar D3$ branes break
supersymmetry and give rise to a metastable positive vacuum energy
state, which is also being felt along the three spatial
dimensions. This acts as a source for
inflation~\cite{Linde1,Frey1} on the MSSM branes. We have depicted
this framework in Fig.~1.

\begin{figure}
\centerline{
\epsfxsize=150pt
\epsfbox{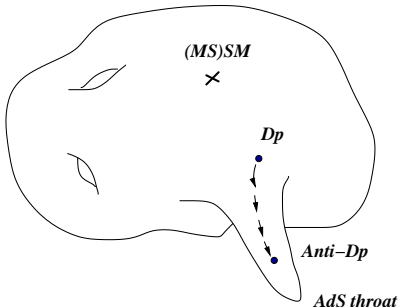}
}
\vskip1ex

\caption{An illustration of a manifold which has singular points
with non-trivial fluxes, and which yields the adS geometry. The
cross denotes the point in moduli space where MSSM branes are
fixed, while there is a $\bar D3$ situated close to the warped
geometry near adS throat, and the $D3$ brane is attracted towards
the $\bar D3$ brane, thereby giving rise to inflation along three
spatial directions, which we assume to be parallel to MSSM branes.
Reheating occurs near the adS throat. The excited modes from
reheating are trapped near the throat. }
\end{figure}

We assume that enough inflation can be obtained along the three
spatial dimensions of the MSSM branes. In the brane anti-brane
scenario inflation ends when the separation becomes close to the
string scale, whence the open string tachyon on the world volume
condenses~\cite{Sen}, resulting in an annihilation of the pair of
branes. Similar situation could arise in our case. The rolling
tachyon couples to the gauge fields living on the brane through
covariant derivatives. The annihilation of branes ultimately gives
rise to a long excited closed string along the inflated
directions~\cite{Maldacena2}. The long closed string decays very
late into lighter closed string modes.  However, the important
point to note here is that only the bulk degrees of freedom near
the adS throat are excited. This is so because the branching ratio
of the closed string decaying into the bulk is still greater
compared to decay into brane degrees of freedom by virtue of phase
space arguments. These modes are actually trapped near the
infrared regime which is energetically more favorable due to low
energy configuration.

This situation is indeed quite similar to the $5$ dimensional adS
Randall-Sundrum model described in Sect. III. There the bulk
quanta were dumped towards the adS horizon to form a black hole.
Formation of a long closed string after the end of inflation also
has a counterpart in field theory. If the inflaton has a global
$U(1)$ charge, it may not decay completely but rather fragments
into lumps known as $Q$-balls~\cite{Kasuya}.

The main point here is that the MSSM branes are not directly
reheated from the decay of the long closed string. They will
certainly feel the resulting effective dark energy, but due to the
warped metric the dark energy is redshifted.

%


\section{Discussion}


During inflation, massless MSSM fields, or the (almost) flat
directions corresponding to certain combinations of squarks,
sleptons and Higgses, will be subject to fluctuations and form
condensates. Like the ordinary inflaton, the condensates will
receive scale-invariant spatial perturbations. Once the inflaton
energy has disappeared into the bulk, the potential terms along
some flat direction will eventually start to dominate the energy
density of the MSSM brane. A potential arises because of
non-renormalizable interactions and the soft supersymmetry
breaking mass terms~(for a review, see \cite{Enqvist}). (In the
cosmological context the Higgs coupling $\mu H_{u}H_{d}$ does not
spoil flatness as $\mu$ is much smaller than the relevant field
amplitudes).

Once the condensate decays, it will reheat the universe with MSSM
degrees of freedom and imprint on the MSSM gas the inflationary
perturbations. The flat direction condensate acts as an MSSM
curvaton~\cite{Sloth,Lyth,Moroi}. The simplest possibility is the
flat direction that consists of the Higgses $H_u$ and $H_d$, which
has been discussed in detail in \cite{Kasuya1}. The reheat
temperature can be estimated to be less than $10^9$ GeV. The
amplitude of the fluctuations along the $H_{u}H_{d}$ direction can
match the observed density perturbations in the CMB radiation and
the spectrum with a spectral tilt very close to 1, with some weak
dependence on the Higgs
potential~\cite{Kasuya1}.

We have argued that a brane cosmology with a warped, infinite
extra dimension allows for the inflaton to decay into the bulk and
for the subsequent redshifting of the decay products. The inflaton
decays efficiently into a continuum of Kaluza Klein modes which
carry non-zero momentum along the extra dimension and move away
from our four dimensional world, taking inflaton energy with them.
In effect, the effective dark energy present on the MSSM brane
will disappear into the bulk and be hidden behind the particle
horizon at the infinity of the 5th dimension.

Although we have sketched a possible inflationary scenario
involving $\bar D3$ and $D3$ branes that annihilate at an adS
throat, a string theoretical model for reheating remains a
challenge. Nevertheless, it seems that at least within brane world
cosmologies it is possible to have the inflaton decay products
disappearing into the bulk so that all matter could have its
origin in the decay of the MSSM condensate rather than in the
inflaton energy density.


\section{Acknowledgements}

K.E. is supported partly by the Academy of Finland grant no. 75065, and
A.M. is a CITA-National fellow and his work is also supported in
part by NSERC (Canada) and by the Fonds de Recherche sur la
Nature et les Technologies du Qu\'ebec.


\end{document}